\documentclass[a4paper,fleqn,usenatbib]{mnras}
\usepackage[T1]{fontenc}
\usepackage{ae,aecompl}
\usepackage{graphicx}	
\usepackage{amssymb}
\usepackage{amsmath}	
\usepackage{times}

\def\eq#1{{Eq.~(\ref{#1})}}

\newcommand{\HI}{H\textsc{i}}

\title[A halo model for neutral hydrogen]{Constraining a halo model for cosmological neutral hydrogen}
\author[Padmanabhan and Refregier]{Hamsa
Padmanabhan\thanks{Electronic address: hamsa.padmanabhan@phys.ethz.ch} and
Alexandre Refregier\thanks{Electronic address: {alexandre.refregier@phys.ethz.ch}
}\\
Institute for Astronomy, ETH Zurich, Wolfgang-Pauli-Strasse 27, CH-8093 Z\"{u}rich, Switzerland}

\date{Accepted ---. Received ---;in original form ---}

\pubyear{2016}

\begin{document}
\label{firstpage}
\pagerange{\pageref{firstpage}--\pageref{lastpage}}
\maketitle

\begin{abstract}
We describe a combined halo model to constrain the distribution of neutral hydrogen (\HI{}) in the post-reionization universe. We combine  constraints from the various probes of \HI{} at different redshifts: the low-redshift 21-cm emission line surveys, intensity mapping experiments at intermediate redshifts, and the Damped Lyman-Alpha (DLA) observations at higher redshifts. We use a Markov Chain Monte Carlo (MCMC) approach to combine the observations and place constraints on the free parameters in the model. Our  best-fit model involves a relation between neutral hydrogen mass $M_{\rm HI}$ and halo mass $M$ with a non-unit slope, and an upper and a lower cutoff.  We find that the model fits all the observables but leads to an underprediction of the bias parameter of DLAs at $z \sim 2.3$. We also find indications of a possible tension between the \HI{} column density distribution and the mass function of \HI{}-selected galaxies at $z\sim 0$. We provide the central values of the parameters of the best-fit model so derived. We also provide a fitting form for the derived evolution of the concentration parameter of  \HI{} in dark matter haloes, and  discuss the implications for the redshift evolution of the \HI{}-halo mass relation.
\end{abstract}

\begin{keywords}
cosmology: observations -- radio lines: galaxies -- galaxies: evolution
\end{keywords}

\section{Introduction}

Studying the evolution of neutral hydrogen (\HI{}) in the post-reionization universe offers key insights into cosmology and galaxy formation. Intensity mapping, in which the \HI{} intensity fluctuations are mapped out without the resolution of individual galaxies, is a novel technique which allows the study of \HI{} on large scales and promises unprecedented constraints on galaxy formation and evolution, cosmology \citep{chang10, masui13, switzer13, santos2015, bull2015} and stringent constraints on models of dark energy \citep[e.g.,][]{chang2008} and modified gravity \citep[e.g.,][]{hall2013}. The 21-cm intensity power spectrum is also an important tracer of the underlying large-scale structure in the post-reionization universe ($z \sim 0-6$), due to the absence of the complicated reionization astrophysics at these epochs. 

The key astrophysical ingredients in the estimation of the 21-cm intensity mapping power spectrum are the density parameter of \HI{}, $\Omega_{\rm HI}$, and its bias $b_{\rm HI}$, which measures how the \HI{} is clustered with respect to the underlying dark matter. Constraints on these parameters come from a variety of astrophysical probes (a detailed summary is available in \citet{hptrcar2015}, hereafter Paper 1). 
At low redshifts, \HI{} gas in galaxies and their environments is primarily studied through blind 21-cm emission line surveys of nearby galaxies like the \HI{} Parkes All Sky Survey \citep[HIPASS;][]{zwaan05} and the Arecibo Fast Legacy ALFA (ALFALFA) Survey \citep{martin10}. These provide measurements of the mass function of \HI{}-selected galaxies, from which the density parameter of \HI{} in galaxies, $\Omega_{\rm HI, gal}$ can be derived. The clustering of the \HI{}-selected galaxies, which constrains the galaxy bias $b_{\rm HI, gal}$, can also be measured from these surveys \citep{martin12}. At intermediate redshifts ($z \sim 0-1$), intensity mapping experiments typically constrain the combination $\Omega_{\rm HI} b_{\rm HI}$ \citep[e.g.,][]{switzer13}.

At higher redshifts, the major reservoirs of \HI{} gas in the post-reionization universe are the Damped Lyman-Alpha (DLA) systems. These systems have very high column densities ($> 10^{20.3}$ cm$^{-2}$) and are found to contain more than 80 \% of the neutral hydrogen at redshifts 2-5 \citep[e.g.,][]{lanzetta1991}. DLAs have been studied through line-of-sight absorption against bright background quasars \citep[][]{noterdaeme12, prochaska09}, as well as direct imaging surveys \citep{fumagalli2014, fumagalli2015}. The study of DLA systems enables the measurements of the column density distribution, $f_{\rm HI}(N_{\rm HI})$ of the DLAs at various redshifts, the incidence $dN/dX$  of the DLAs  per unit absorption distance interval, and the density parameter of \HI{} in DLAs, $\Omega_{\rm DLA}$. Recently, the large-scale bias parameter of DLA systems, $b_{\rm DLA}$ has also been constrained through a cross-correlation study with the Lyman-$\alpha$ forest \citep{fontribera2012}.

Analytical and simulation techniques have typically been used to model the \HI{} radio observations at lower redshifts \citep[e.g.,][]{dave2013, bagla2010, guhasarkar2012, kim2016, villaescusa2015} and the DLA observations at higher redshifts \citep{fumagalli2011, bird2014, barnes2014} separately. Modelling the the 21-cm based observables requires knowledge of the \HI{} - halo mass relation, $M_{\rm HI} (M)$ which is combined with the radial distribution of \HI{} (the \HI{} density profile, $\rho_{\rm HI} (r)$) to derive the DLA parameters. In \citet[][hereafter Paper 2]{hptrcar2016}, we combined the analytical approaches in the literature towards a consistent picture of \HI{} evolution across redshifts. We fitted the available data both from the 21-cm based as well as from the DLA-based observations by modifying the 21-cm based approach in the literature to also include the DLA profile. We found that the free parameters can be fixed by fitting to the combined set of 21-cm and DLA observations, however, a model that is consistent with the low-redshift observations leads to a prediction for the bias parameter of the DLAs that is lower than the observed value. On the other hand, a  model that is consistent with all observations requires very rapid evolution of the nature of \HI{} bearing host dark matter haloes between redshifts 0-2. Hence, it was found that bringing together the low-and high-redshift observations, while not having been attempted in the literature before, has interesting consequences for the evolution of the free parameters involved in the modelling of \HI{}.

In this paper, we revisit our previous modelling of \HI{} to combine all the available data within the fully statistical framework of a Markov Chain Monte Carlo (MCMC) analysis, using a six free parameter halo model which is a generalization of the \HI{}-halo mass relation used in previous literature. The model also generalizes the  radial distribution of \HI{} to take into account the effective redshift evolution of the concentration of \HI{} in the dark matter haloes. We use a larger dataset compared to the previous work, including also the \HI{} mass function at $z \sim 0$ \citep{zwaan05}.  We find that the resulting combination of data, together with the generalized model, enables fairly strong constraints on the free parameters. This also enables us to explore the correlations between the parameters and their posterior distributions. We again find that a model that is reasonably consistent with all observations underpredicts the measured DLA bias at $z \sim 2.3$. We also find indications of a possible tension between the \HI{} mass function and the column density distribution at $z \sim 0$. The evolution of the free parameters in the model has important physical implications for the power spectrum of \HI{} 21-cm intensity fluctuations, as also for the connections between \HI{} and galaxies.  

The paper is organized as follows. In the next section, we describe the details of the halo model, focusing on the free parameters involved and their physical implications for the \HI{} distribution in the dark matter haloes.  We also review the procedure for obtaining the DLA- and \HI{} based quantities from the analytical model.  In Sec. \ref{sec:data}, we briefly describe the details of the data sets used in the analysis. In the next section, Sec. \ref{sec:mcmc}, we outline the MCMC analysis used for the combination of the data, and obtain the best-fitting model from the analysis. We compare the model predictions to the observations where available, and describe the model predictions at higher redshifts. We summarize the physical implications of our findings and the future outlook in a concluding section (Sec. \ref{sec:summary}).

\section{A halo model for neutral hydrogen}
\label{sec:model}
In this section, we describe the details of the modelling of \HI{} and the physical implications of the free parameters involved in the model. 
Given the dark matter halo mass distribution, a prescription is required to populate \HI{} in the haloes, which involves specifying both (i) the HI mass- halo mass relation $M_{\rm HI} (M)$,  and (ii) the radial density profile $\rho_{\rm HI} (r)$ which describes the distribution of \HI{} within the halo.

The  $M_{\rm HI} - M$  relation is modelled as:
\begin{eqnarray}
M_{\rm HI} (M) &=& \alpha f_{H,c} M \left(\frac{M}{10^{11} h^{-1} M_{\odot}}\right)^{\beta} \exp\left[-\left(\frac{v_{c0}}{v_c(M)}\right)^3\right] \nonumber \\
&\times &\exp\left[-\left(\frac{v_{c} (M)}{v_{c1}}\right)^3\right]
\end{eqnarray}
which is a generalization of the HI-halo mass relations used previously in the literature \citep[][and Paper 2]{barnes2014}. The relation involves the following quantities:
\begin{enumerate}
\item  the multiplicative constant $\alpha$, the fraction of \HI{} relative to the cosmic $f_{H,c}$
\item the excess (logarithmic) slope, $\beta$, of the $M_{\rm HI} - M$ relation, taken to be zero in previous studies (so that the overall power of $M$ was unity), the relation is pivoted at $10^{11} h^{-1} M_{\odot}$;
\item the two cutoffs in the virial velocity $v_c (M)$ of the halo; $v_{c0}$ (the lower cutoff) and $v_{c1}$ (the upper cutoff), used previously in Paper 2; the lower cutoff alone has been used in DLA-based analyses such as \citet{barnes2014}. 
\end{enumerate}

Each of the factors in the above $M_{\rm HI} - M$ relation has a physical implication. The multiplicative constant, $\alpha$, denotes the fraction of hydrogen relative to cosmic and is directly related to the redshift evolution of \HI{} and its depletion due to star-formation and feedback processes. From observational and simulation studies of the neutral hydrogen evolution, this value is expected to be fairly constant or slowly evolving over redshifts \citep[e.g.,][]{crighton2015, dave2013}. 

The value of $\beta$ represents the correction to unity of the logarithmic slope of the \HI{}-halo mass relation. It is seen from the results of abundance matching \citep[e.g.,][]{papastergis2013} that the observationally favoured value of $\beta$ is less than zero, such that the power of the halo mass in the \HI{}-halo mass relation  is less than unity. 

The two cutoffs, $v_{c0}$ (the lower cutoff) and $v_{c1}$ (the upper cutoff) determine, at different redshifts, the range of dark matter halo masses that contribute chiefly as hosts of \HI{} in galaxies. Besides being of direct relevance in determining the bias factor of neutral hydrogen (as seen in e.g., Paper 2), the two cutoffs are also related to physical processes that influence the masses of the dark matter haloes that are able to host \HI{}. It is found as a result of simulations \citep{pontzen2008}, that halos with virial velocities of less than 30 km/s and greater than 200 km/s preferentially do not host neutral hydrogen. However, these cutoffs, especially the  upper cutoff $v_{c1}$, are also directly linked to $\beta$, which provides a measure of the effective weights given to the range of dark matter halo masses that host \HI{}. Hence, these two cutoffs are also important in the contexts of galaxy formation and feedback processes leading to the generation and destruction of \HI{}. 

Together with the above \HI{}-halo mass relation, we also use a radial density profile, $\rho_{\rm HI} (r)$ to describe the distribution of \HI{} in the dark matter halo. This is modelled following \citet{maller2004}, the form being motivated by the simulation constraints on cooling in multi-phase gas in haloes:
\begin{equation}
\rho_{\rm HI} (r) = \frac{\rho_0 r_s^3}{(r + 0.75 r_s) (r+r_s)^2}
\label{rhodefnew}
\end{equation}
The form is an altered Navarro-Frenk-White \citep[NFW;][]{navarro1997} profile, where  $r_s$ is the scale radius of the halo, defined through $r_s = R_v(M)/c(M,z)$, with $R_v (M)$ being the virial radius.
The \HI{} concentration parameter, $c(M,z)$ is given by:
\begin{equation}
 c(M,z) = c_{\rm HI} \left(\frac{M}{10^{11} M_{\odot}} \right)^{-0.109} \frac{4}{(1+z)^{\gamma}}.
\end{equation}

In the above relation, $c_{\rm HI}$ is a free parameter,  analogous to the dark matter halo concentration $c_0 = 3.4$ in, e.g., \citet{maccio2007}. 
This form of the profile is found to reproduce well the observed properties of high-redshift DLAs,\footnote{Simulation studies and observations of gas-rich galaxies \citep[e.g.,][]{obreschkow2009, wang2014} suggest the low redshift \HI{} surface density profiles to be well represented by exponential disks. Due to the additional complexity needed to model the projection angles associated with the disk, as well as a possible bulge component, we leave the exploration of variations to the low-redshift \HI{} profile to future work. The present form of the profile also bridges the low-and high redshift analyses without introducing additional degrees of freedom into the model.} though it is generally observed \citep[][Paper 2]{barnes2010, barnes2014} that the value of $c_{\rm HI}$ is higher than $c_0$. In other words,  the \HI{} gas is more tightly concentrated than the dark matter. The value of $\gamma$ in the above relation was taken to be unity in previous analyses, including the case for dark matter alone. In the present analysis, it signifies the extent of the evolution correction due to the redshift evolution of the parameter $c_{\rm HI}$, i.e. the value $\gamma = 1$ translates into \textit{no} redshift evolution of the concentration parameter $c_{\rm HI}$.

The value of $\rho_0$ in \eq{rhodefnew} is determined by normalization:
\begin{equation}
 \int_0^{R_v(M)} 4 \pi r^2 \rho_{\rm HI}(r) dr = M_{\rm HI} (M)
\end{equation} 
with the total mass of \HI{} in the halo. 
Hence, all the free parameters in the model are involved in the description of the \HI{} profile.

Once the $M_{\rm HI} - M$ relation and the form of the \HI{} profile are known, the various quantities related to the \HI{} observations can be derived analytically (a detailed description of the method is available in Paper 2):
\begin{enumerate}
\item The column density as a function of the impact parameter $s$ of a line-of-sight through a DLA system:
\begin{equation}
 N_{\rm HI}(s) = \frac{2}{m_H} \int_0^{\sqrt{R_v(M)^2 - s^2}} \rho_{\rm HI} (r = \sqrt{s^2 + l^2}) \ dl 
 \label{coldenssnew}
\end{equation} 
where $m_H$ is the proton mass, and the cross section is defined by $\sigma_{\rm DLA}(M) = \pi s_*^2$ where $s_*$ is the impact parameter corresponding to  the column density reaching the DLA limit: $N_{\rm HI} (s_*) = 10^{20.3} {\rm cm}^{-2}$.

\item The incidence, $dN/dX$ and the bias of the DLAs, $b_{\rm DLA}$, both of which can be derived from the cross section:
\begin{equation}
 \frac{dN}{dX} = \frac{c}{H_0} \int_0^{\infty} n(M,z) \sigma_{\rm DLA}(M,z) \ dM \ ,
 \label{dndxdef}
\end{equation} 
\begin{equation}
 b_{\rm DLA} (z) =  \frac{\int_{0}^{\infty} dM n (M,z) b(M,z) \sigma_{\rm DLA} (M,z)}{\int_{0}^{\infty} dM n (M,z) \sigma_{\rm DLA} (M,z)} \ .
\end{equation} 
\item The column density distribution $f_{\rm HI}(N_{\rm HI}, z)$, given by:
\begin{equation}
 f(N_{\rm HI}, z) = \frac{c}{H_0} \int_0^{\infty} n(M,z) \left|\frac{d \sigma}{d N_{\rm HI}} (M,z) \right| \ dM 
 \label{coldensdef}
\end{equation} 
where $d \sigma/d N_{\rm HI} =  2 \pi \ s \ ds/d N_{\rm HI}$, $H_0$ is the present-day Hubble parameter, and $N_{\rm HI} (s)$ is defined as in \eq{coldenssnew}.
 In all the above expressions, $n(M,z)$ denotes the dark matter mass function. As in the previous analysis (Paper 2), we take this to be given by the Sheth-Tormen (ST) mass function \citep{sheth2002}.

\item The density parameter of the DLAs, $\Omega_{\rm DLA}$, given by integration of the column density distribution:
\begin{equation}
 \Omega_{\rm DLA}(N_{\rm HI}, z)  = \frac{m_H H_0}{c \rho_{c,0}} \int_{10^{20.3}}^{\infty} f_{\rm HI}(N_{\rm HI}, z) N_{\rm HI} d N_{\rm HI}
 \label{omdlanew}
\end{equation}
where $\rho_{c,0}$ is the present-day critical density of the universe. 
\item The \HI{} mass function (measured, for example, in surveys of \HI{}-selected galaxies)\footnote{Note that if the $M_{\rm HI}(M)$ is not a monotonic function, the contributions from the two sides of the turnover point are added when computing the term $|dM/dM_{\rm HI}|$. In practice, it is found that the contribution from masses higher than the maximum value of $M_{\rm HI} (M)$ is negligible, so we assume a sharp cutoff (instead of exponential) at the maximum value of $M_{\rm HI} (M)$ for the computation.} :
\begin{equation}
\phi(M_{\rm HI}, z) = n (M,z)  \left|\frac{d M}{d M_{\rm HI}} (M,z) \right| 
\end{equation}
\item The density parameter and the bias of the HI:
\begin{equation}
 \Omega_{\rm HI} (z) = \frac{1}{\rho_{c,0}} \int_0^{\infty} n(M, z) M_{\rm HI} (M,z) dM
 \label{omegaHInew}
\end{equation} 
\begin{equation}
b_{\rm HI} (z) = \frac{\int_{0}^{\infty} dM n(M,z) b (M,z) M_{\rm HI} (M,z)}{\int_{0}^{\infty} dM n(M,z) M_{\rm HI} (M,z)}
\label{biasHInew}
\end{equation}
with the dark matter halo bias $b(M,z)$ following, e.g., \citet{scoccimarro2001}.
\end{enumerate}

We can use the available data to constrain the free parameters in the model above. The available data are summarized in the next section and include the column density distribution at redshifts 0, 1 and 2.3, the large-scale bias of \HI{} selected galaxies from ALFALFA at $z \sim 0$ \footnote{Here, as in Paper 2, we make the approximation that the selection function of the galaxies can be modelled by a combination of mass-weighting and the integral cutoffs, and hence use Eq. \ref{biasHInew} to compare with the data. Note that for simplicity, we do not consider the effects of subhalos that host satellite galaxies in the present work, since our analysis here is limited to the large scale \HI{} observables.}, the bias of DLA systems estimated at $z \sim 2.3$, and the DLA incidence $dN/dX$ from other observations at these redshifts. We use a Markov Chain Monte Carlo approach to constrain the free parameters $c_{\rm HI}, \alpha, \beta, \gamma, v_{c0}$ and $v_{c1}$ from the observations, as summarized in Sec. \ref{sec:mcmc}.

\section{Data}
\label{sec:data}
In this section, we provide a brief summary of the data at different redshifts used to fit to the model and estimate the best-fitting values of the free parameters (a detailed summary of the data related to \HI{} observables across redshifts are in Papers 1 and 2):

(i) The column density distribution of DLAs at redshifts 0 from the WHISP survey \citep{zwaan2005a}, from redshift 1 \citep{rao06}, and the observations at redshift 2.3 \citep{noterdaeme12},

(ii) the DLA incidence \citep{zwaan2005a, rao06, braun2012, zafar2013} at the above redshifts, the value of $\Omega_{\rm DLA}$ at $z \sim 2.3$ from \citet{zafar2013},

(iii) the \HI{} mass function at $z \sim 0$ from the HIPASS survey \citep{zwaan05},

(iv) the product $\Omega_{\rm HI} b_{\rm HI}$ at $z \sim 1$ from the results of intensity mapping \citep{switzer13}, and

(v) the clustering of the DLA systems at $z \sim 2.3$ \citep{fontribera2012}.

Of these data, (i), (ii), (iv) and (v) were also used in the previous analysis comparing different models of the HI and DLA observables (Paper 2). The \HI{} mass function, added in to the present statistical analysis provides additional constraints on the parameters $\alpha$, $\beta$, $v_{c0}$ and $v_{c1}$ which appear in the \HI{}-halo mass relation. 

Figures \ref{fig:redshift0}, \ref{fig:redshift1} and \ref{fig:redshift23} show the compilation of this data at the  redshifts 0, 1 and 2.3 respectively.

\section{Combining constraints: MCMC analysis}
\label{sec:mcmc}
We use a Markov Chain Monte Carlo (MCMC) analysis with the \textsc{CosmoHammer} routine \citep{akeret2013} to explore the available parameter space, combine the constraints and sample the likelihood to find the best-fitting values and errors on the free parameters. The cosmological parameters used are consistent with the earlier analysis in Paper 2: a flat $\Lambda$ CDM cosmology with $\Omega_m =0.281$, $\Omega_{\Lambda} = 0.719$, $\Omega_b  = 0.0462$, $\sigma_8 = 0.81$, $n_s = 0.963$, and $h = 0.71$.

\subsection{Parameters and priors}
We parametrize the model by the six free parameters: $c_{\rm HI}, \alpha, \beta, \gamma$, $v_{c0}$ and $v_{c1}$. The model assumes flat priors for each of the parameters, with the ranges for the priors indicated in the second column of Table \ref{table:initial}.

\begin{table}
\centering
\caption{Summary of the free parameters,  their flat prior ranges, the best-fitting values and their uncertainties obtained from the model.}
\begin{tabular}{llllll}
\hline
Parameter     &  Range for prior & Best fit &  Error (1$\sigma$)  \\
\hline \\
$c_{\rm HI}$ & [20, 400]  &  113.80 & 14.03\\
$\alpha$        & [0.05, 0.5] & 0.17  & 0.02 \\
log $v_{c0}$  (km/s)   & [1.30, 1.90] & 1.57 & 0.03\\
log $v_{c1}$ (km/s)    & [2.1, 6.5] &  4.39 & 0.75 \\
$\beta$         & [-1,3] &  -0.55 & 0.12 \\
$\gamma$        & [-0.9,2] & 0.22  & 0.67\\
 \hline
\end{tabular}
\label{table:initial}
\end{table}

\subsection{Likelihood}
The likelihood is defined by:
\begin{equation}
\mathcal{L} = \exp\left(-\chi^2/2\right)
\end{equation}
where the $\chi^2$ is calculated by:
\begin{equation}
\chi^2 = \sum_i \frac{(g_{i, obs} - g_i)^2}{\sigma_i^2}
\end{equation}
In the above equation, the $g_i$'s are the model predictions for the different quantities, the $g_{i, obs}$ are the observed values, and the $\sigma_i$'s are the errors on the observations  which are assumed independent. {{We note that the observations of the column density distribution and the \HI{} mass function at $z \sim 0$ provide only the statistical errors. To obtain an indicative measure of the systematic effects on the derived parameter values, we:

(a) use the systematic uncertainties on the parameters of the best-fit Schechter form of the HIPASS mass function at $z \sim 0$ \citep{zwaan05} and add these in quadrature to the statistical uncertainties; the resulting error band is shown in grey in the second panel of Fig. \ref{fig:redshift0}, and

(b) perform the analysis separately with the WHISP \citep{zwaan2005a} and the \citet{braun2012} measurements of the column density distribution at $z \sim 0$, in order to provide a measure of the effect of the systematic uncertainty on the derived best-fit parameters. We describe the results of fitting to the WHISP sample in the main text, the comparison to the \citet{braun2012} data is provided in the Appendix.
}} 

\subsection{Results}

\begin{figure}
\begin{center}
\includegraphics[scale=0.4, width = \columnwidth]{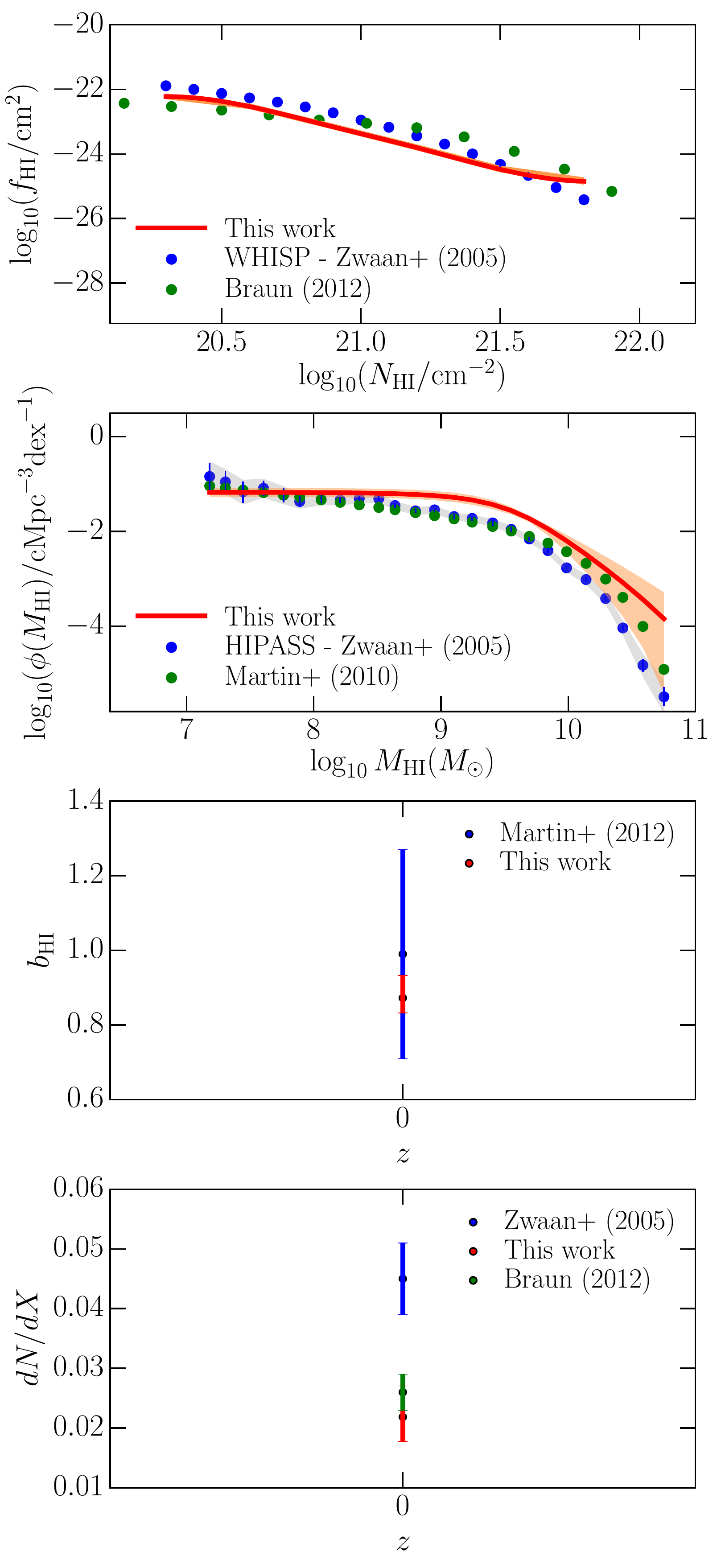} 
\end{center}
\caption{The data and analysis for $z \sim  0$. The data sets for the column density distribution $f_{\rm HI}$ (top panel) include (i) WHISP \citep{zwaan2005a}, which was used in the fitting, and (ii) the results from  \citet{braun2012}, from the measurements of \HI{} in local galaxies. The mass function $\phi(M_{\rm HI})$  of \HI{} selected galaxies (second panel) is from the HIPASS survey \citep{zwaan05}.  {The grey band indicates the overall observational error estimated by adding the systematic and statistical uncertainties in quadrature. The mass function from the ALFALFA survey \citep{martin10} is also shown for comparison.} The bias function of \HI{}-selected galaxies from ALFALFA \citep{martin12} is shown in the third  panel. The values of the incidence, $dN/dX$ (fourth panel) are shown from both the WHISP and the \citet{braun2012} surveys. }
\label{fig:redshift0}
\end{figure}

\begin{figure}
\begin{center}
\includegraphics[scale=0.4, width = \columnwidth]{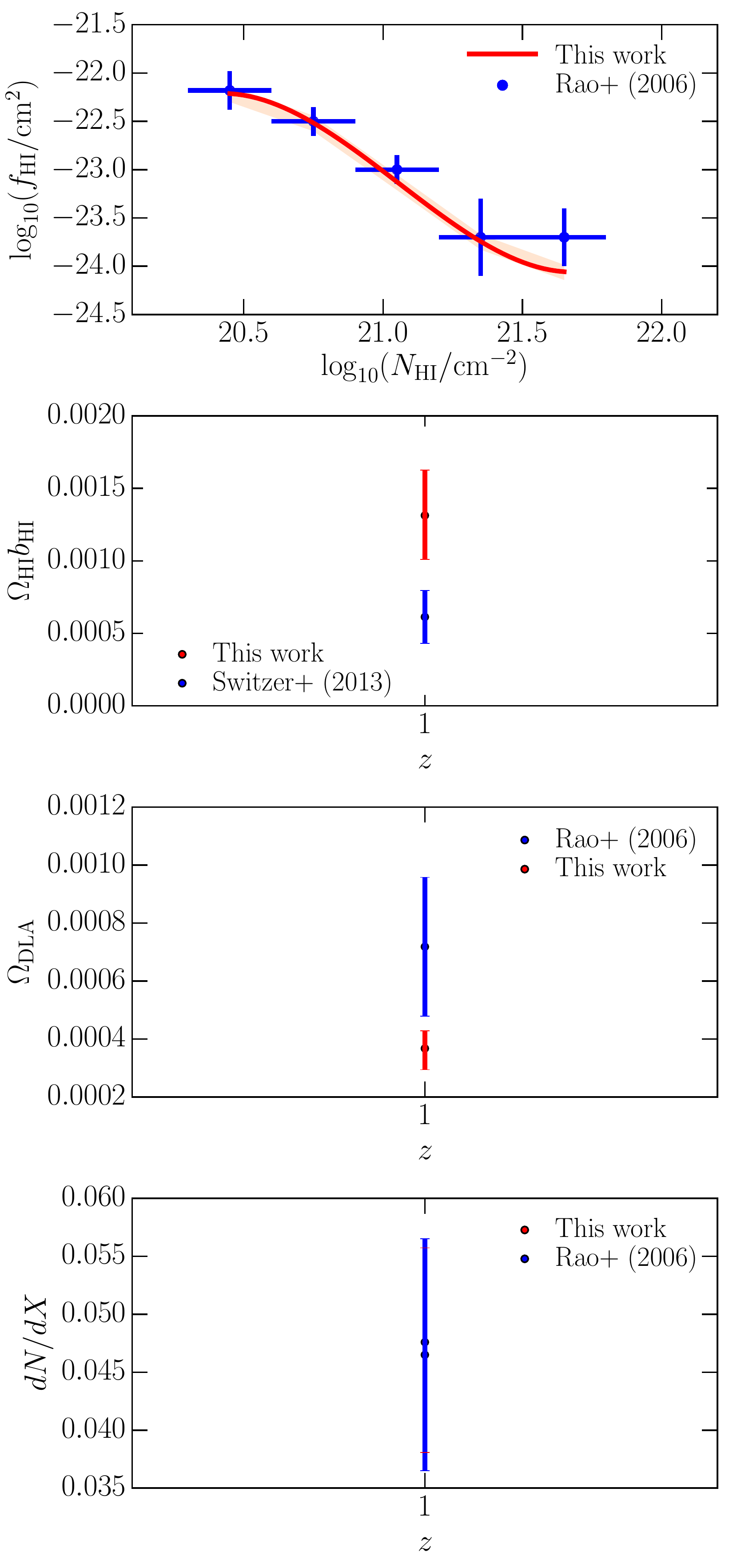} 
\end{center}
\caption{The data and analysis for $z \sim 1$. The data points with error bars include the column density distribution $f_{\rm HI}$ \citep[top panel;][]{rao06}. The second panel shows the product $\Omega_{\rm HI} b_{\rm HI}$ from the results of intensity mapping \citep{switzer13}. The  derived parameters: the DLA incidence $dN/dX$ (third panel) and the DLA density parameter $\Omega_{\rm DLA}$ (fourth panel) are also shown.}
\label{fig:redshift1}
\end{figure}

\begin{figure}
\begin{center}
\includegraphics[scale=0.6, width = \columnwidth]{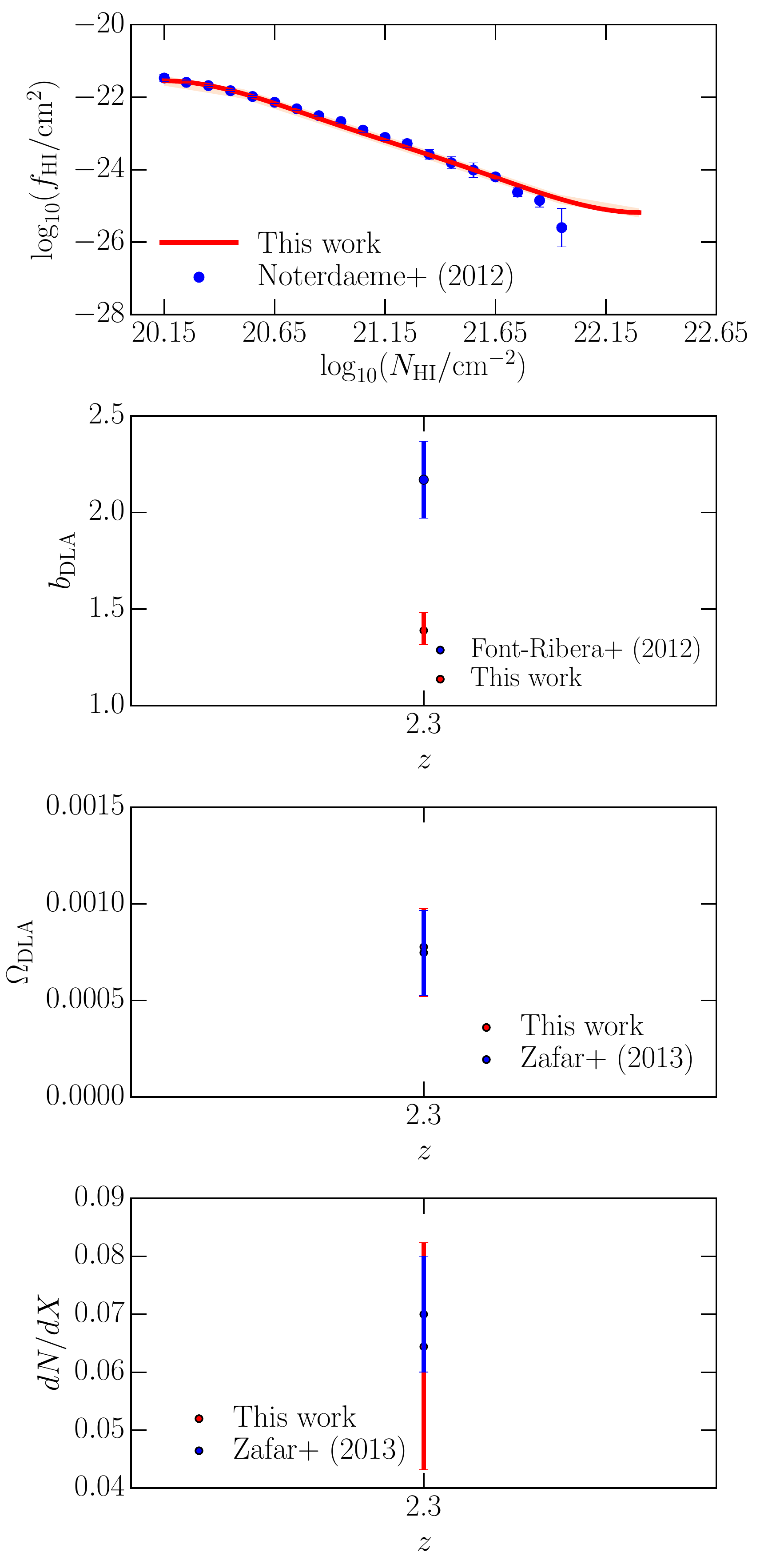} 
\end{center}
\caption{The data and analysis for $z \sim 2.3$. The data points include  the column density distribution $f_{\rm HI}$ (top panel) from the results of \citet{noterdaeme12},  and  the clustering of DLAs, $b_{\rm DLA}$ from the analysis of \citet[][second panel]{fontribera2012}. Also shown are the density parameter $\Omega_{\rm DLA}$ (third panel)  and  incidence $dN/dX$ (fourth panel) from \citet{zafar2013}.}
\label{fig:redshift23}
\end{figure}

We sample the likelihood using 100 chains each of 10 random walkers for each of the 6 parameters, making a total of 6000 chains. \footnote{We find the likelihood to be adequately converged for the above number of iterations. For computational efficiency, we restrict the fitting of the column density distribution to column densities upto $N_{\rm HI} \sim 22$ cm$^{-2}$, however, since the function is monotonic with little or no structure, we find that the best-fitting model is an adequate match at higher column densities as well.} The best-fitting values and the uncertainties so derived for the parameters are in the last two columns of Table \ref{table:initial}. For the quoted best-fit values and errors, we use the mean and $1 \sigma$ uncertainties of the values from the MCMC analysis.

\begin{figure*}
\begin{center}
\includegraphics[width = 2\columnwidth]{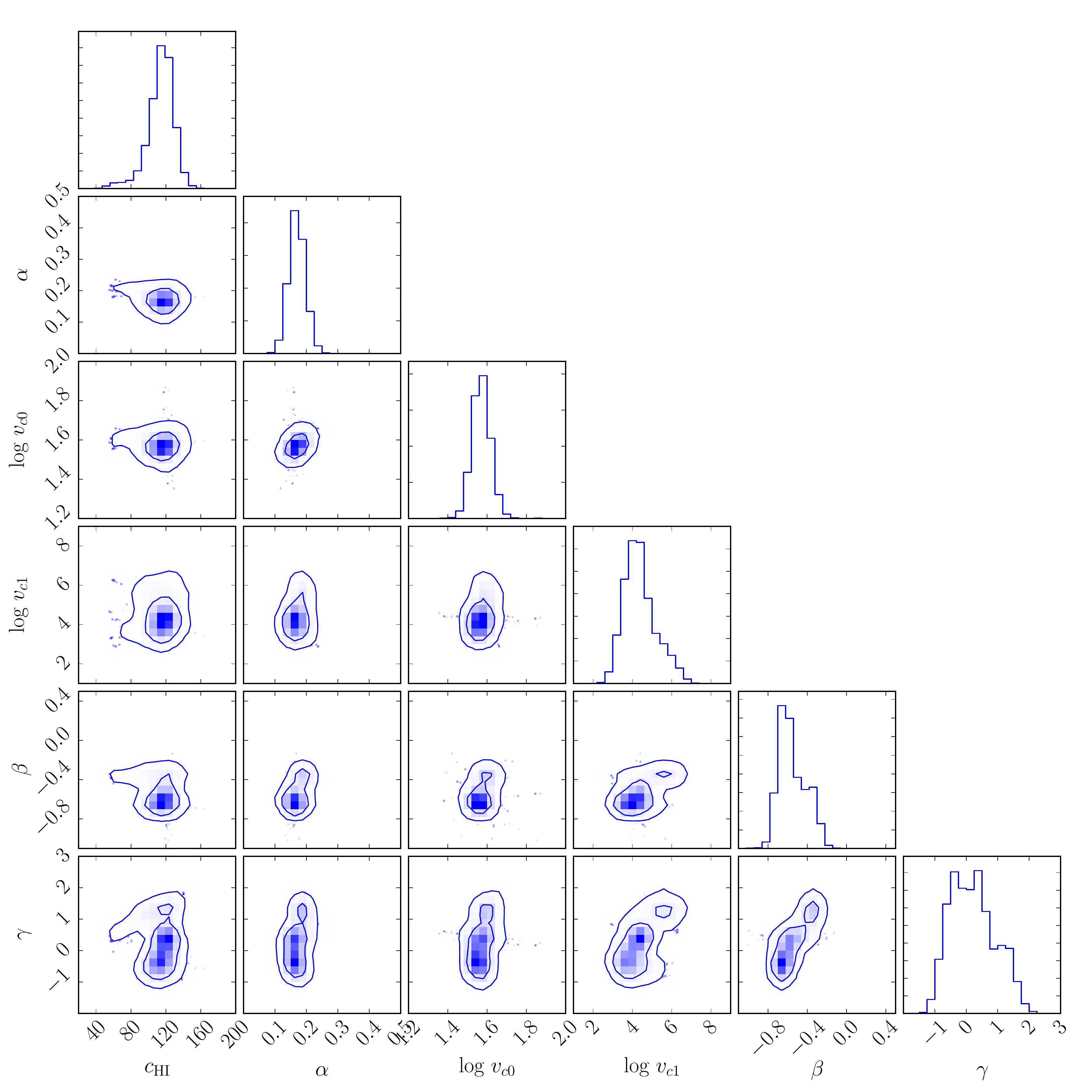} 
\end{center}
\caption{The likelihood space of the six parameters, $c_{\rm HI}, \alpha,  v_{c0},  v_{c1}, \beta$, and $\gamma$, as sampled using the Monte Carlo analysis, using the \textsc{Cornerplot} routine (Foreman-Mackey  2016). The posterior distributions of the parameters are indicated in the diagonal  panels. Contours show the 68 and 95 percent confidence levels.}
\label{fig:cornerplot}
\end{figure*}

\begin{figure*}
\begin{center}
\includegraphics[scale=0.6, width = \textwidth]{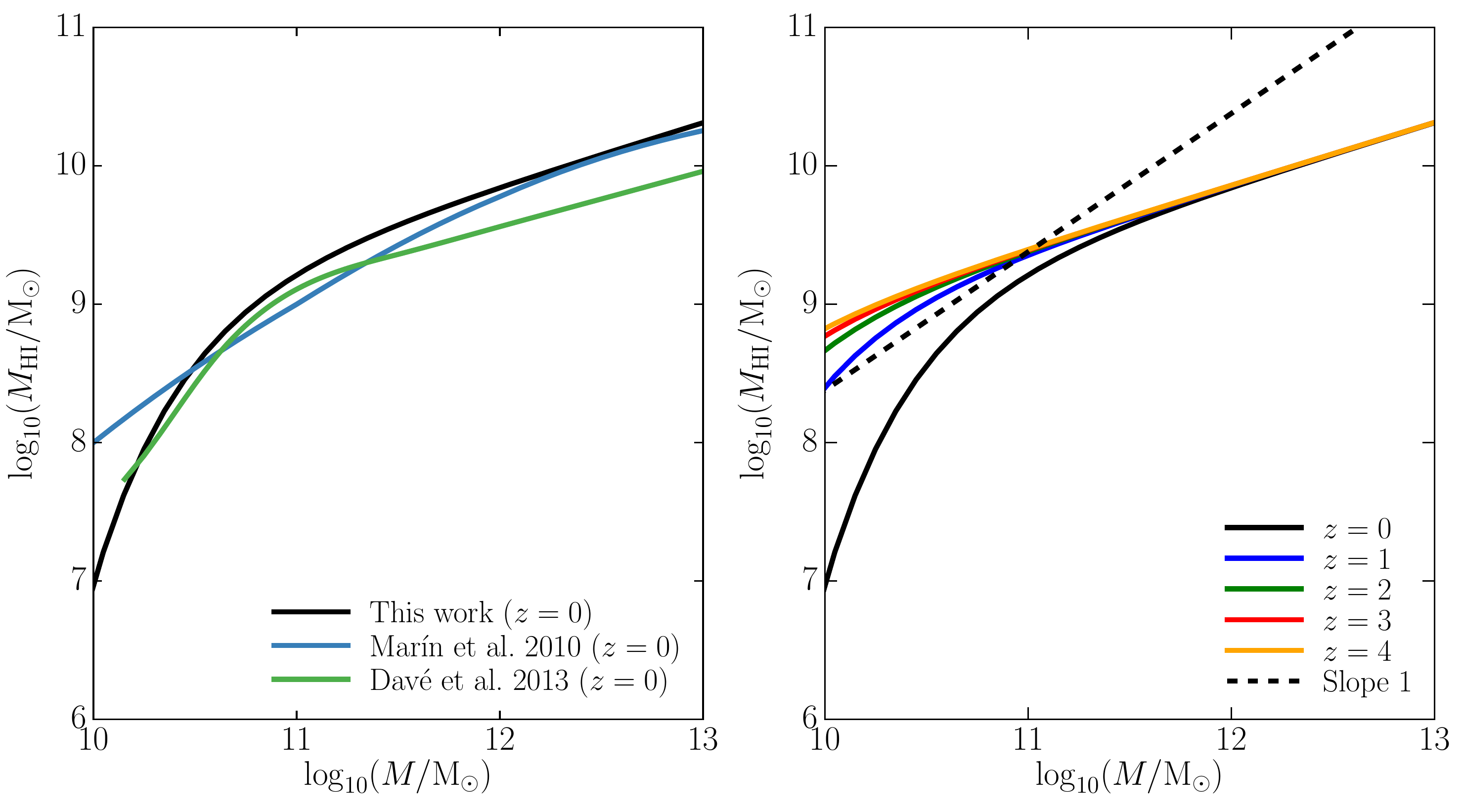}
\end{center}
\caption{\textit{Left panel:} the best-fitting $M_{\rm HI} - M$ relation at $z \sim 0$, along with the results of simulations in the literature \citep{marin2010, dave2013}. \textit{Right panel:} The evolution  of the best-fitting $M_{\rm HI} - M$ relation across redshifts (plots of the relation at five redshifts are shown). A plot of the relation with logarithmic slope unity, i.e. $M_{\rm HI} \propto M$, is also shown for comparison (dashed line).}
\label{fig:mhim}
\end{figure*}

A comparison of the best-fit model to the data is provided in the  panels of Figs. \ref{fig:redshift0}, \ref{fig:redshift1} and \ref{fig:redshift23}, for the redshifts 0, 1 and 2.3 respectively.  The bands and error bars show the observables predicted by the best-fitting model with the mean values for the six parameters. The uncertainties are determined by propagating the errors in the parameters in quadrature. 
Most of the observables are fairly well fit by the model predictions, apart from an under prediction of $b_{\rm DLA}$ at $z \sim 2.3$ (also seen in the previous study, Paper 2). There are also indications of a possible tension between the mass function of \HI{}-selected galaxies and the column density distribution at $z \sim 0$. Note that, at low redshifts,  the systematic uncertainties in the reported measurements of the column density distribution and the mass function may be significant compared to the statistical uncertainties. To provide an indicative measure of the systematic effects, we have plotted multiple measurements where available, for example, the \citet{braun2012} and the \citet{zwaan2005a} measurements of the column density distribution, and the \citet{zwaan05} and \citet{martin10} measurements of the mass function at $z \sim 0$. {To explore the possibility of the observed tension being alleviated by the systematics in the measurements at $z \sim 0$, especially in the case of $f_{\rm HI}$, we repeat the analysis with the \citet{braun2012} data (in place of the WHISP data) in the Appendix. We find, as a result of this analysis, that the systematics in the data may not fully alleviate the observed tension.}

The posterior probabilities from the MCMC analysis, and the parameter space are shown in Fig. \ref{fig:cornerplot}. 
The resultant form of $M_{\rm HI} (M)$ at $z = 0$ is shown in the left panel of Fig. \ref{fig:mhim}, along with the results of other studies in the literature \citep{marin2010, dave2013}. The evolution of the $M_{\rm HI} (M)$ relation for redshifts 0-4 is shown in the right panel of Fig. \ref{fig:mhim}, along with the relation for $M_{\rm HI} \propto M$ (shown by the dashed line). It can be seen that the $M_{\rm HI} - M$ relation derived from the best-fitting model has a shallower slope in all cases, illustrating the effect of negative values for $\beta$.

There is no redshift evolution in the present $M_{\rm HI} (M)$, apart from that coming from the evolution of the virial velocity to halo mass relation. A possible evolution in the parameter $\beta$ may be needed to match the observed clustering bias of DLAs at $z \sim 2.3$, which we explore in greater detail in future work. The evolution of the $M_{\rm HI} - M$ relation can be constrained by current and future intensity mapping experiments which measure the large-scale power spectrum $P_{\rm HI} (k)$. There are also studies that infer the mean host halo mass of DLA systems at high redshift through clustering and cross-correlation studies  \citep{bouche2005, fontribera2012}. Cross-correlations of the 21-cm intensity maps with a large sample of optical galaxies \citep[e.g.][]{chang10, masui13, switzer13} can potentially provide powerful constraints on the \HI{}-halo mass relation.

\begin{figure}
\begin{center}
\includegraphics[scale=0.6, width = \columnwidth]{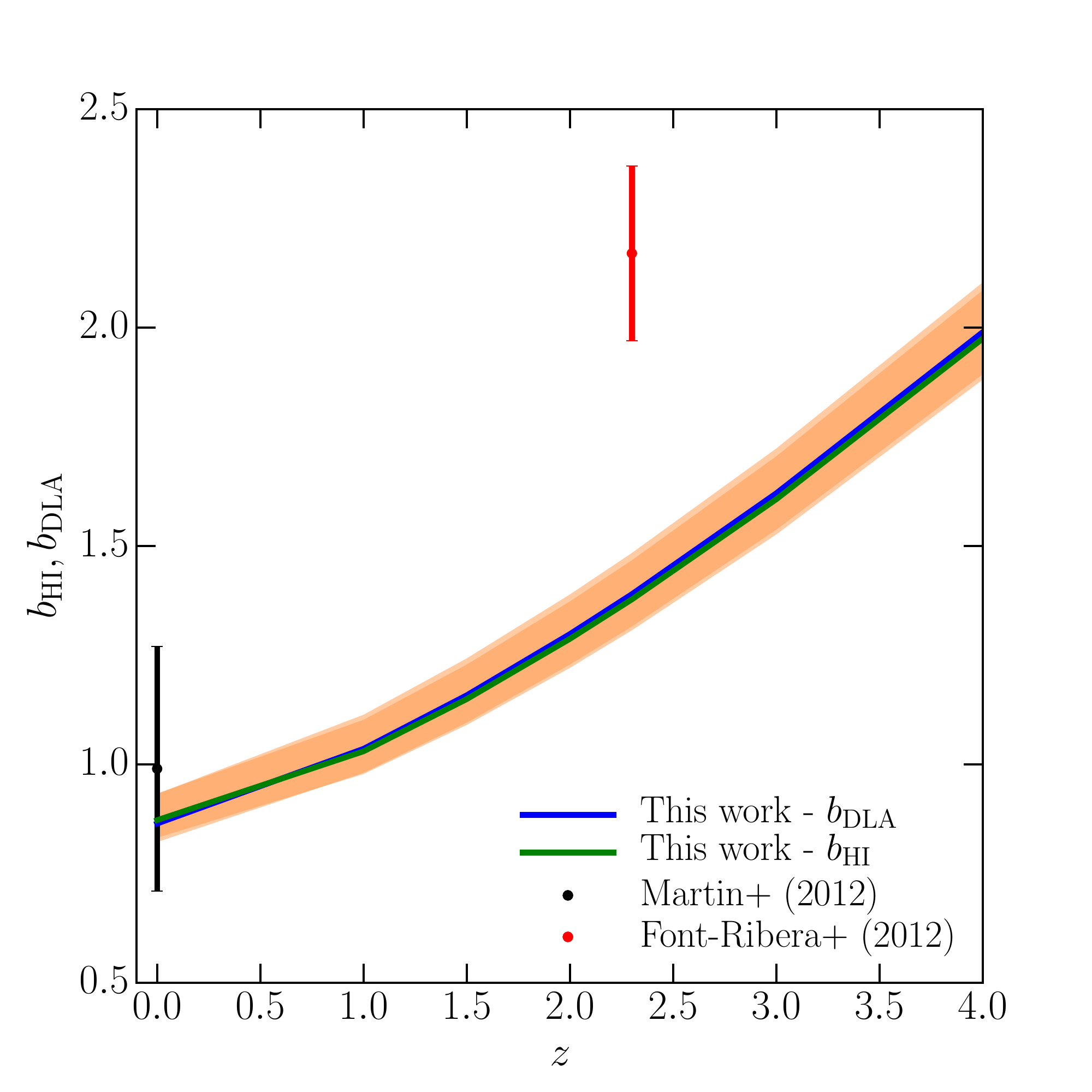}
\end{center}
\caption{The evolution of the bias, both for \HI{} as well as for DLAs, as a function of redshift in the best-fitting model. The error band indicates the range allowed by the uncertainties in the model parameters. The data include the bias of \HI{}-selected galaxies at $z \sim 0$ \citep{martin12} and the bias parameter of the DLAs at $z \sim 2.3$ \citep{fontribera2012}.}
\label{fig:bias}
\end{figure}

\begin{figure}
\begin{center}
\includegraphics[scale=0.6, width = \columnwidth]{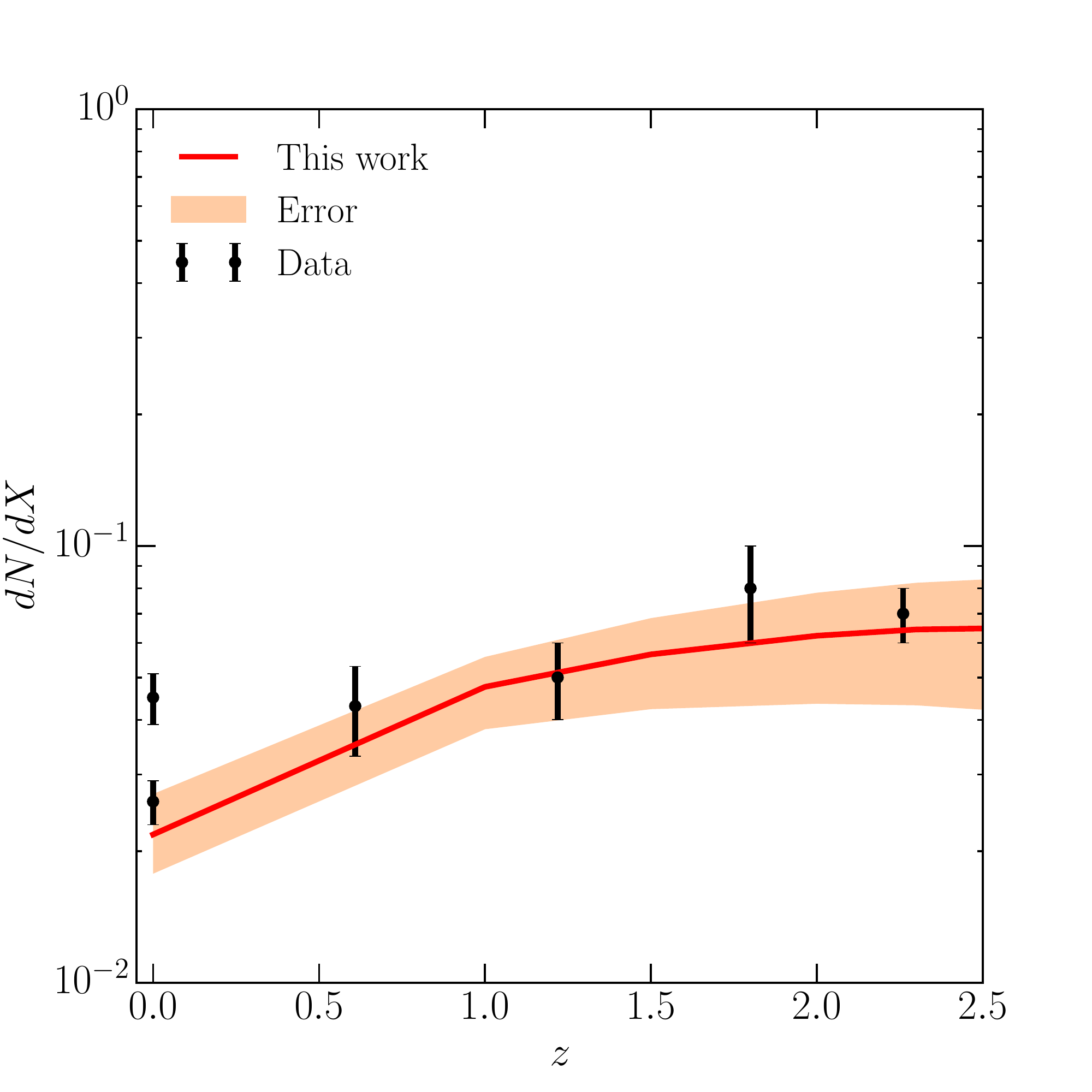}
\end{center}
\caption{The evolution of the incidence, $dN/dX$ across redshifts with the best-fitting model, with the error band indicating the range of allowed values from the model uncertainties. The data points are the results of (i) at redshift 0: \citet{zwaan2005a} and  \citet{braun2012}, (ii) at redshifts $\sim 1$:  \citet{rao06}; (iii) at higher redshifts: \citet{zafar2013}.}
\label{fig:dndx}
\end{figure}

Finally, we plot the model predictions for the evolution of the bias, both for \HI{} and DLAs, and the incidence $dN/dX$ as a function of redshift. The errors on these are again determined by propagating in quadrature the uncertainties in the free parameters. These predictions are shown along with the data points in Figs. \ref{fig:bias} and \ref{fig:dndx} respectively. The predictions for $b_{\rm HI}$ at intermediate and high redshifts are important in the context of current and upcoming intensity mapping experiments.

 We thus find that the additional input of the \HI{} mass function at low-$z$, constrains the value of $\beta$ to less than zero, thereby making the logarithmic slope of the \HI{}-halo mass relation lower than unity. This slope is directly related also to the high-mass cutoff in the \HI{}-halo mass relation (the value of $v_{c1}$). A smaller slope allows a higher value of $v_{c1}$ to be consistent with the measurements, since it results in a lower weight being given to the high mass end of the \HI{}-halo mass relation. We thus see that the introduction of the parameter $\beta$ is essential for obtaining a good fit to the low-redshift \HI{} mass function and contributes towards constraining the overall "shape" of the \HI{}-halo mass relation. 

\begin{figure}
\begin{center}
\includegraphics[scale=0.6, width = \columnwidth]{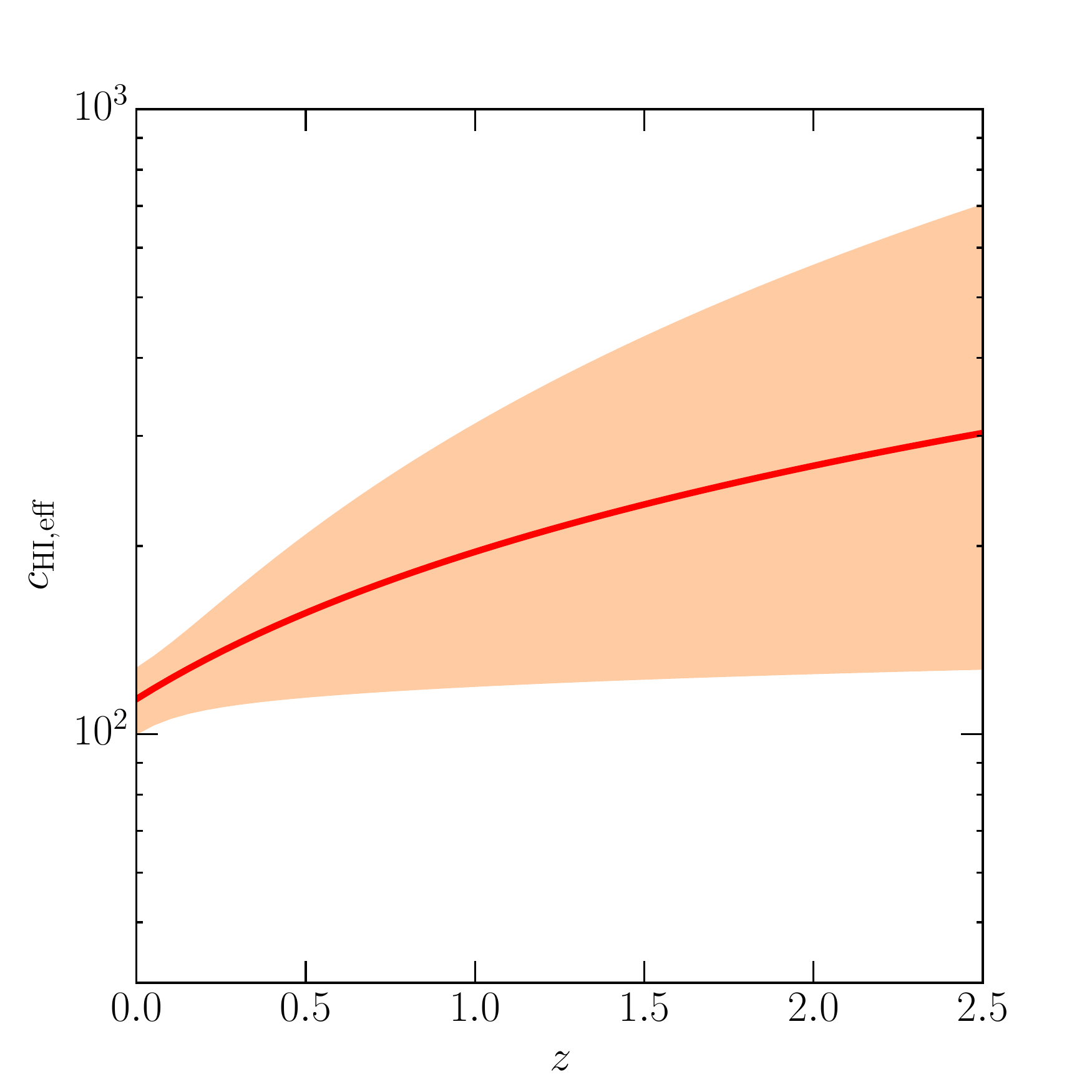}
\end{center}
\caption{The evolution of the "effective" concentration parameter of the \HI{}, $c_{\rm HI, eff}$ as a function of redshift. The model uncertainties are indicated by the error band. The effective evolution of the concentration parameter scales approximately as $(1 + z) ^{0.5}$.}
\label{fig:chieff}
\end{figure}

  The parameter $\gamma$, used in the redshift evolution of the \HI{} profile,  is also found to converge to a value less than unity. 
This quantifies the redshift evolution of the \HI{} concentration parameter compared to that expected from dark matter alone. If we define an "effective" concentration parameter $c_{\rm HI, eff} = c_{\rm HI} (1 + z) ^{1 - \gamma}$,  we thus find evidence for a non-trivial redshift evolution of the $c_{\rm HI, eff}$, with the remaining evolution in the profile being identical to that of the dark matter. This evolution of the effective concentration of \HI{} is shown in Fig. \ref{fig:chieff} with its error band, which indicates a mild increase of the best-fitting \HI{} concentration with increasing redshift.

\section{Conclusions}
\label{sec:summary}
In this paper, we have carried out a detailed statistical study of the  evolution of \HI{} across redshifts by utilizing all the available data, both from the 21-cm based as well as the DLA-based observables.
We build upon the results of our previous fitting to the observations (Paper 2) by generalizing the form of $M_{\rm HI} (M)$ and the \HI{} profile from older analyses, thus introducing extra degrees of freedom. We thus explore how well we can use the data available at present to develop a halo-model based approach towards analyzing \HI{}. The model free parameters are constrained using a full MCMC  analysis so that the posterior distributions and the correlations between parameters can be effectively explored. We find that a model which is reasonably consistent with all available data underpredicts the value of DLA bias at $z \sim 2.3$, as shown in Figs. \ref{fig:redshift23} and \ref{fig:bias}. This was also observed with the previous fitting analysis (Paper 2). The model predictions are nevertheless consistent with the results of DLA imaging studies, which favour the association of high-redshift DLAs with faint dwarf galaxies \citep{cooke2015, fumagalli2014, fumagalli2015}, and the results of hydrodynamical simulations \citep[e.g.,][]{rahmati2014, dave2013}. 

The introduction of the new parameters $\beta$ and $\gamma$ constrains the shape of the \HI{}-halo mass relation and also the form and evolution of the \HI{} density profile. Constraining $\beta < 0$ results in a smaller weight being given to the higher range of host halo masses, thus allowing the upper cutoff $v_{c1}$ to be higher than for the case with $\beta = 0$. It is also seen from the results of abundance matching \citep{papastergis2013} that a flatter $M_{\rm HI} - M$ relation is favoured by the observations of ALFALFA \HI{} galaxies. However, there are indications of a possible tension between the \HI{} column density distribution and the mass function of \HI{}-selected galaxies at $z \sim 0$. This tension may not be fully alleviated by the systematic effects in the measurements of the column density distribution and the mass function at $z \sim 0$. It is possible that the form of the profile at low redshifts needs to be modified in order to fit the observed column density distribution. We hope to consider other forms of the profile such as an exponential surface density profile, as well as a possible variation of the form of the profile across redshifts in future work.

With the introduction of the parameter $\gamma$, the redshift evolution of the concentration $c_{\rm HI}$ can be effectively explored, and it can be seen that a mild increase in \HI{} concentration with redshift is favoured by the best-fitting model. 

In future work, it would be interesting to extend the results of this halo model framework to include small-scale clustering, measured from surveys of \HI{} selected galaxies and DLA systems.This serves to provide additional constraints on the profile parameters, as well as their evolution across redshifts. This extension would also enable  comparison to the results of hydrodynamical simulations, and thus shed light on the connections between \HI{} and galaxy evolution over cosmic time. Finally, such a combined halo model would have  important consequences for constraining the expected signal in the  \HI{}  power spectrum, to be measured with current and future intensity mapping experiments.

\section*{Acknowledgements}

We thank Tirthankar Roy Choudhury,  Adam Amara, Phil Bull, Girish Kulkarni, Supratik Pal, Sebastian Seehars and Ali Rahmati for useful discussions and comments on the manuscript, and Joel Akeret for help and support with the \textsc{CosmoHammer} package. We thank the anonymous referee for a detailed and  helpful report that improved the content and presentation. The research of HP is supported by the Tomalla Foundation.

\bibliographystyle{mnras} 

\bibliography{mybib}

\bsp

\appendix

\section{Impact of systematic uncertainties on the fitted parameters}
{
In this appendix, we provide an indicative measure of the impact of systematic uncertainties in the measurement of the column density distribution at $z \sim 0$ by repeating the analysis in the main text for the \citet{braun2012} data. The resulting corner plot is shown by the green curves in Fig. \ref{fig:cornerboth} and is overplotted on the corresponding corner plot in the main text (shown in blue). The comparison to the analysis in the main text is summarized in the first four columns of Table \ref{table:systematics}.

\begin{table*}
\centering
\caption{Comparison of the best-fitting values and their uncertainties obtained from (a) fitting to the $z \sim 0$ WHISP column density distribution,  (b) fitting to the $z \sim 0$ Braun (2012) column density distribution, and (c) fitting to a combined dataset with systematic errors assumed to be the differences in the measurements of the mass function and the column density distribution at $z \sim 0$.}
\begin{tabular}{llllllll}
\hline
Parameter     &  Best fit (WHISP) &  Error (1$\sigma$; WHISP)  &  Best fit (Braun) &  Error (1$\sigma$; Braun)  &  Best fit (combined) &  Error (1$\sigma$; combined)  \\
\hline \\
$c_{\rm HI}$ &   113.80 & 14.03  &   128.05  & 12.63 & 123.79 & 15.67\\
$\alpha$        & 0.17  & 0.02  &   0.16 &  0.01 & 0.16 &  0.02\\
log $v_{c0}$  (km/s)   & 1.57 & 0.03  & 1.61   & 0.03 & 1.61 & 0.05\\
log $v_{c1}$ (km/s)     &  4.39 & 0.75   & 4.93   & 0.68 & 4.64 & 0.72\\
$\beta$         &  -0.55 & 0.12  & -0.66   & 0.06  & -0.63 & 0.07\\
$\gamma$         & 0.22  & 0.67  &  0.55  & 0.20 & 0.58 & 0.37\\
 \hline
\end{tabular}
\label{table:systematics}
\end{table*}

\begin{figure*}
\begin{center}
\includegraphics[width = 2\columnwidth]{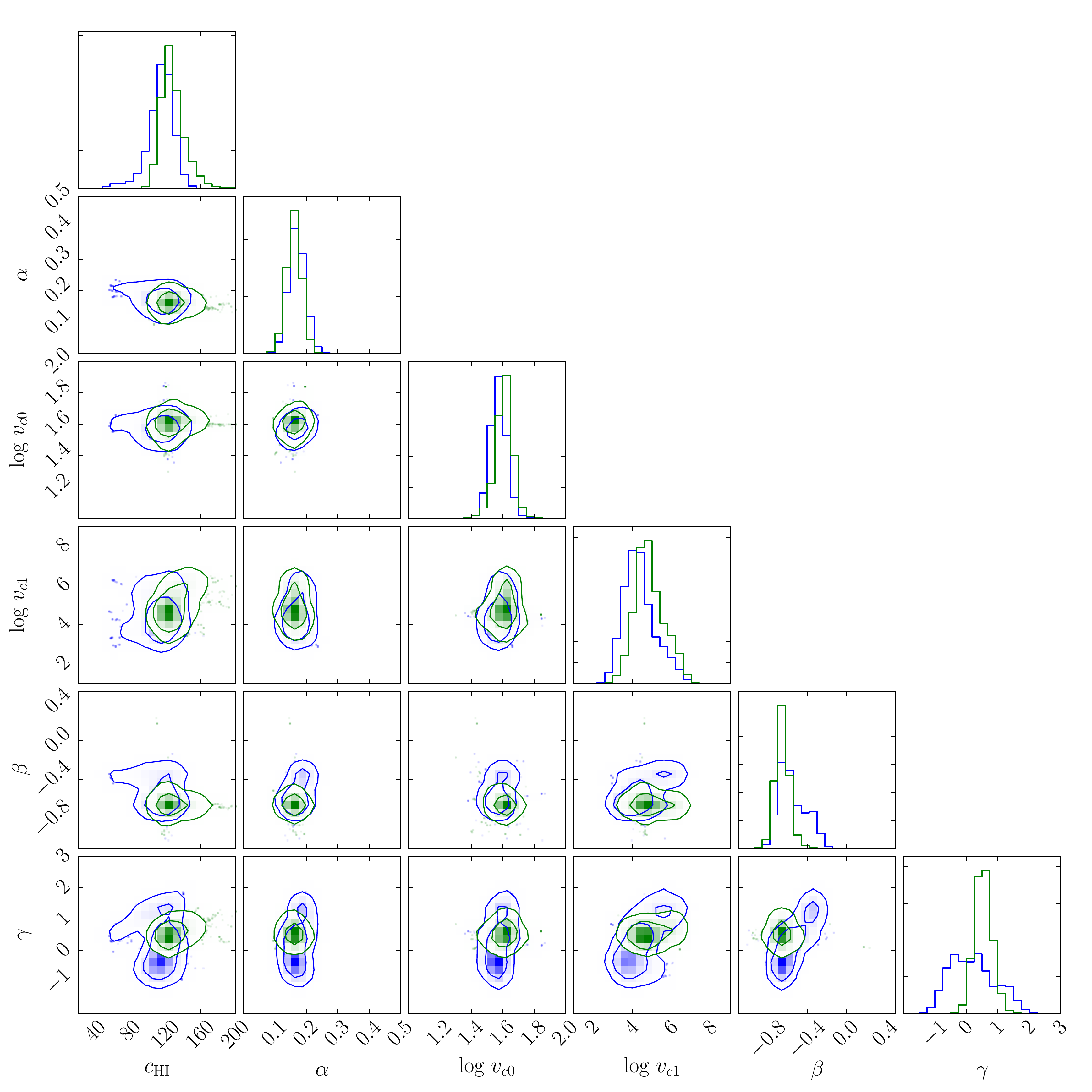} 
\end{center}
\caption{ Same as Fig. \ref{fig:cornerplot}, but now repeated for the column density distribution from \citet{braun2012}. The analysis is in green, and overplotted on the original corner plot (blue).}
\label{fig:cornerboth}
\end{figure*}

An indicative measure of the systematic effects can also be provided by considering the difference between two observational datasets to provide an estimate of systematic errors. We consider the effect of using the difference between the two column density distribution measurements (from the WHISP and \citet{braun2012}), and also the two mass function measurements (from HIPASS; \citet{zwaan05} and ALFALFA; \citet{martin10}) as measures of the systematic uncertainties in the observables. The systematic uncertainty in each case is then added to the statistical uncertainty in quadrature. Upon repeating the analysis, we find the best-fit parameters given by  the fifth and sixth columns of Table \ref{table:systematics}. The best-fit parameter values for the three analyses are consistent within $1\sigma$, indicating that systematic effects may not be completely responsible for the observed tension between the $z \sim 0$ column density and the \HI{} mass function. 
}

\label{lastpage}
\end{document}